\documentclass[letterpaper, 10 pt, conference]{ieeeconf}  % Comment this line out if you need a4paper

\IEEEoverridecommandlockouts                              % This command is only needed if 
\usepackage{cite}
\usepackage{amsmath,amssymb,amsfonts}
\usepackage{algorithmic}
\usepackage{graphicx}
\usepackage{textcomp}
\usepackage{xcolor}

\usepackage{multicol}
\usepackage{wrapfig}
\usepackage{subcaption}
\usepackage{amsmath}
\usepackage[ruled,vlined,linesnumbered]{algorithm2e}

\let\labelindent\relax
\usepackage{enumitem}

\def\BibTeX{{\rm B\kern-.05em{\sc i\kern-.025em b}\kern-.08em
    T\kern-.1667em\lower.7ex\hbox{E}\kern-.125emX}}
\begin{document}

\title{\LARGE \bf
Object Packing and Scheduling for Sequential 3D Printing:\\ a Linear Arithmetic Model and a CEGAR-inspired Optimal Solver
}

\newtheorem{definition}{Definition}
\newtheorem{proposition}{Proposition}
\newtheorem{corollary}{Corollary}

%\title{A Small 3D Printed Robotic Arm for Teaching Industry 4.0 and Robotic Engineering
%\thanks{This work is supported by the Faculty of Information Technology, Czech Technical University in Prague and by GA\v{C}R - the Czech Science Foundation, grant registration number 22-31346S.}
%}

%\author{\IEEEauthorblockN{Pavel Surynek}
%\IEEEauthorblockA{\textit{Faculty of Information Technology, Czech Technical University in Prague} \\
%Th\'{a}kurova 9, 160 00 Praha 6, Czechia \\
%\tt\small pavel.surynek@fit.cvut.cz, ORCID: 0000-0001-7200-0542
%}}

\author{Pavel Surynek$^{1}$ and Vojt\v{e}ch Bubn\'{i}k$^2$ and Luk\'{a}\v{s} Mat\v{e}na$^2$ and Petr Kubi\v{s}$^2$% <-this % stops a space
\thanks{$^{1}$Faculty of Information Technology, Czech Technical University in Prague,
        Th\'{a}kurova 9, 160 00 Praha 6, Czechia
        {\tt\small pavel.surynek@fit.cvut.cz}}       
\thanks{$^{2}$Prusa Research,
        Partyz\'{a}nsk\'{a} 188/7a, 170 00 Praha 7 - Hole\v{s}ovice, Czechia
        {\tt\small \{vojtech.bubnik, lukas.matena, petr.kubis\}@prusa3d.cz}}
% <-this % stops a space        
}

\maketitle

\begin{abstract}

We address the problem of object arrangement and scheduling for sequential 3D printing. Unlike the standard 3D printing, where all objects are printed slice by slice at once, in sequential 3D printing, objects are completed one after other. In the sequential case, it is necessary to ensure that the moving parts of the printer do not collide with previously printed objects. We look at the sequential printing problem from the perspective of combinatorial optimization. We propose to express the problem as a linear arithmetic formula, which is then solved using a solver for satisfiability modulo theories (SMT). However, we do not solve the formula expressing the problem of object arrangement and scheduling directly, but we have proposed a technique inspired by counterexample guided abstraction refinement (CEGAR), which turned out to be a key innovation to efficiency.

%V clanku se zabyvame otazkou arrangementu a schedulingu objektu pro sekvencni 3D tisk. Na rozdil od stadarniho, kdy jsou tisteny po vrstvach najednou vsechny objekty, v sekvencnim 3D tisku jsou objekty dokoncovany jeden po druhem. Pritom je treba zajistit, aby nedoslo ke kolizi pohyblivych casti tiskarny s drive natistenymi objekty. Na ulohu sekvencniho tisku pohlizime z hlediska kombinatoricke optimalizace. Navrhujeme vyjadrid problem pomoci formule linearni aritmetiky, kterou nasledne resime pomoci resice pro satisfiability modulo theories (SMT). Formuli vyjadrujici problem rozmistovani a rozvrhovani objektu ale neresime primo, ale navrhli jsme techniku inspirovanou counterexample guided abstraction refinement (CEGAR), coz se ukazalo jako klicove pro efektivitu.

%We show that the use of CEGAR-style refinements in solving model of the sequential 3D printing problem represents a significant innovation.
%Ukazujeme, ze vyuziti refinements ve stylu CEGAR v reseni sekvencniho 3D tisku predstavuje vyznamnou inovaci.

\end{abstract}

%\begin{IEEEkeywords}
%robotic arm, 3D-printing, robotics engineering, robotics teaching, robot control
%\end{IEEEkeywords}

\section{Introduction}

Additive manufacturing, i.e. 3D printing, is an increasingly important alternative to traditional manufacturing processes. At the same time, 3D printing, due to its nature, is very close to robotics and to the techniques of artificial intelligence and combinatorial optimization that are used in robotics, since the 3D printer itself can be viewed as a special robot \cite{DBLP:journals/cad/GaoZRRCWWSZZ15}.

A standard Cartesian fused deposition modeling (FDM) 3D printer creates objects on a rectangular printing plate (usually heated) by gradually drawing individual slices of printed objects, where these slices are very thin, approximately tenths of a millimeter. Printing is performed using a print head with an extruder, which applies material through a narrow nozzle. The movement of the print head is ensured by a printer mechanism that allows the head to move in all $x$, $y$, and $z$ coordinates.

One of the important tasks of combinatorial optimization in 3D printing is the arrangement of printed objects on the printing plate so that the space of the place is used effectively \cite{multi-objective-packing2014,DBLP:conf/ki/EdelkampW15}.

\begin{figure*}[t]
    \centering
    \begin{subfigure}{0.33\textwidth}
       \includegraphics[trim={0.5cm 0.5cm 0.5cm 0.5cm},clip,width=1.0\textwidth]{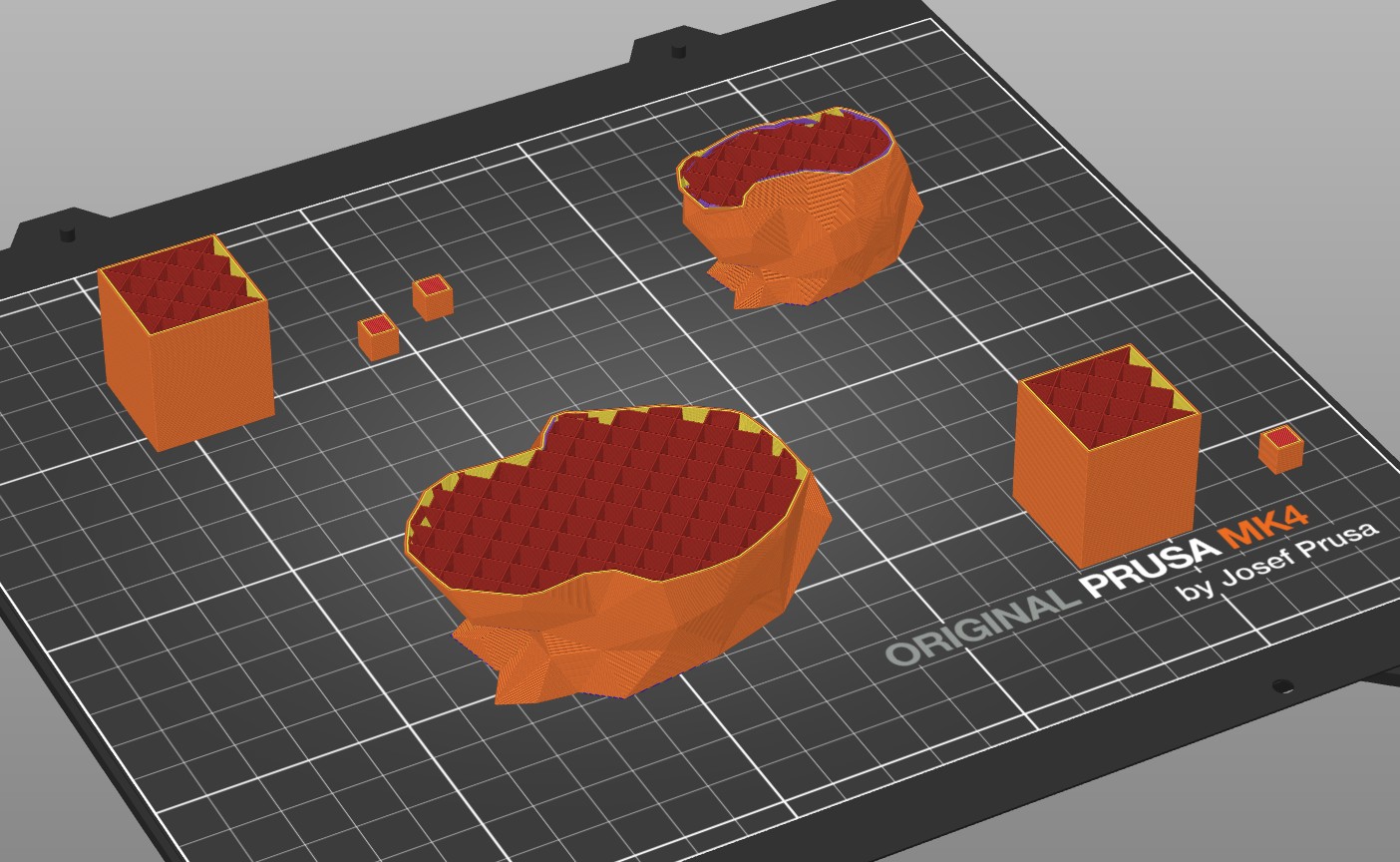}
    \end{subfigure}
    \begin{subfigure}{0.31\textwidth}
       \includegraphics[trim={0.5cm 0.5cm 0.5cm 0.5cm},clip,width=1.0\textwidth]{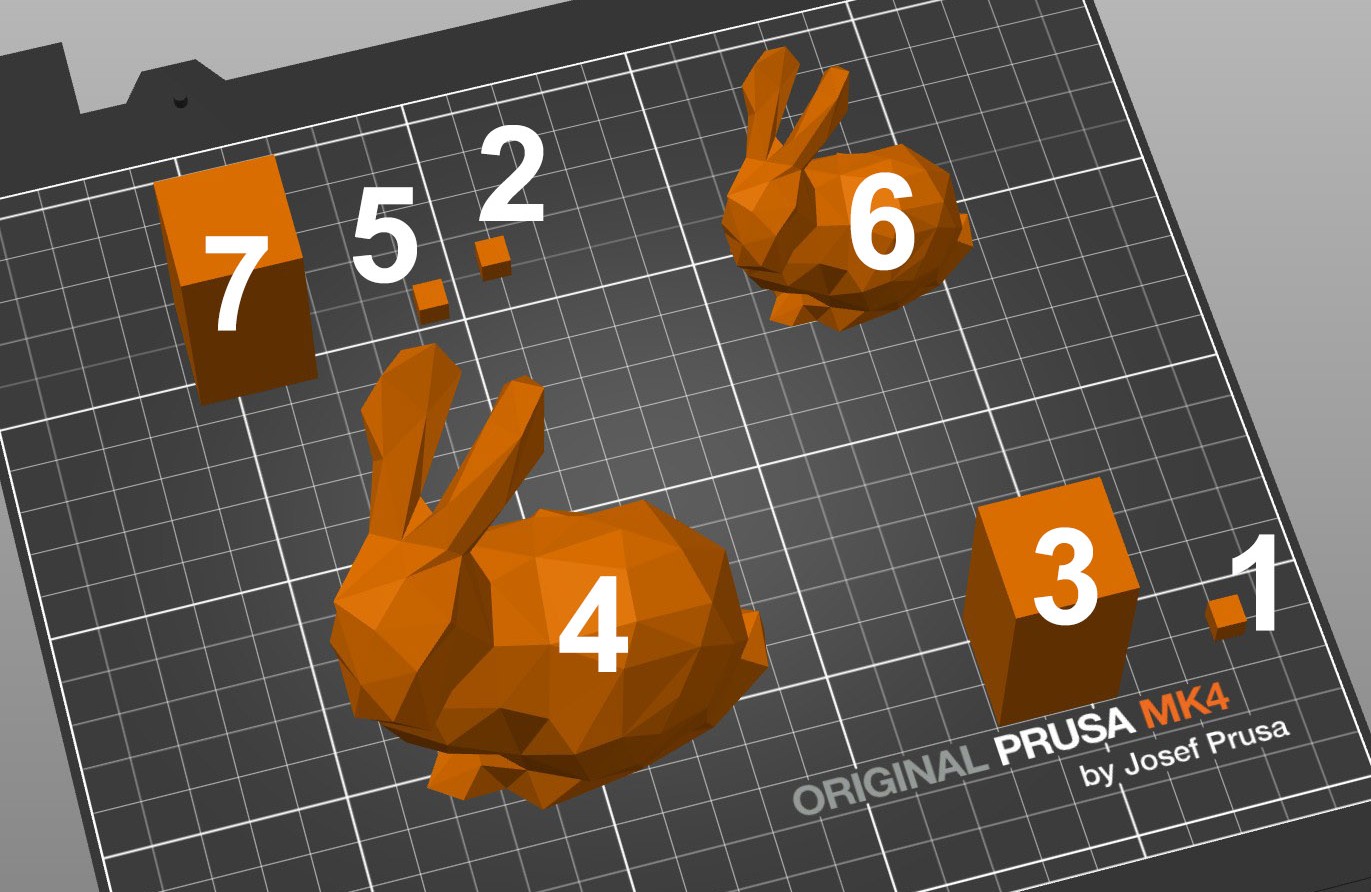}
    \end{subfigure}
    \begin{subfigure}{0.345\textwidth}
       \includegraphics[trim={0.5cm 0.5cm 0.5cm 0.5cm},clip,width=1.0\textwidth]{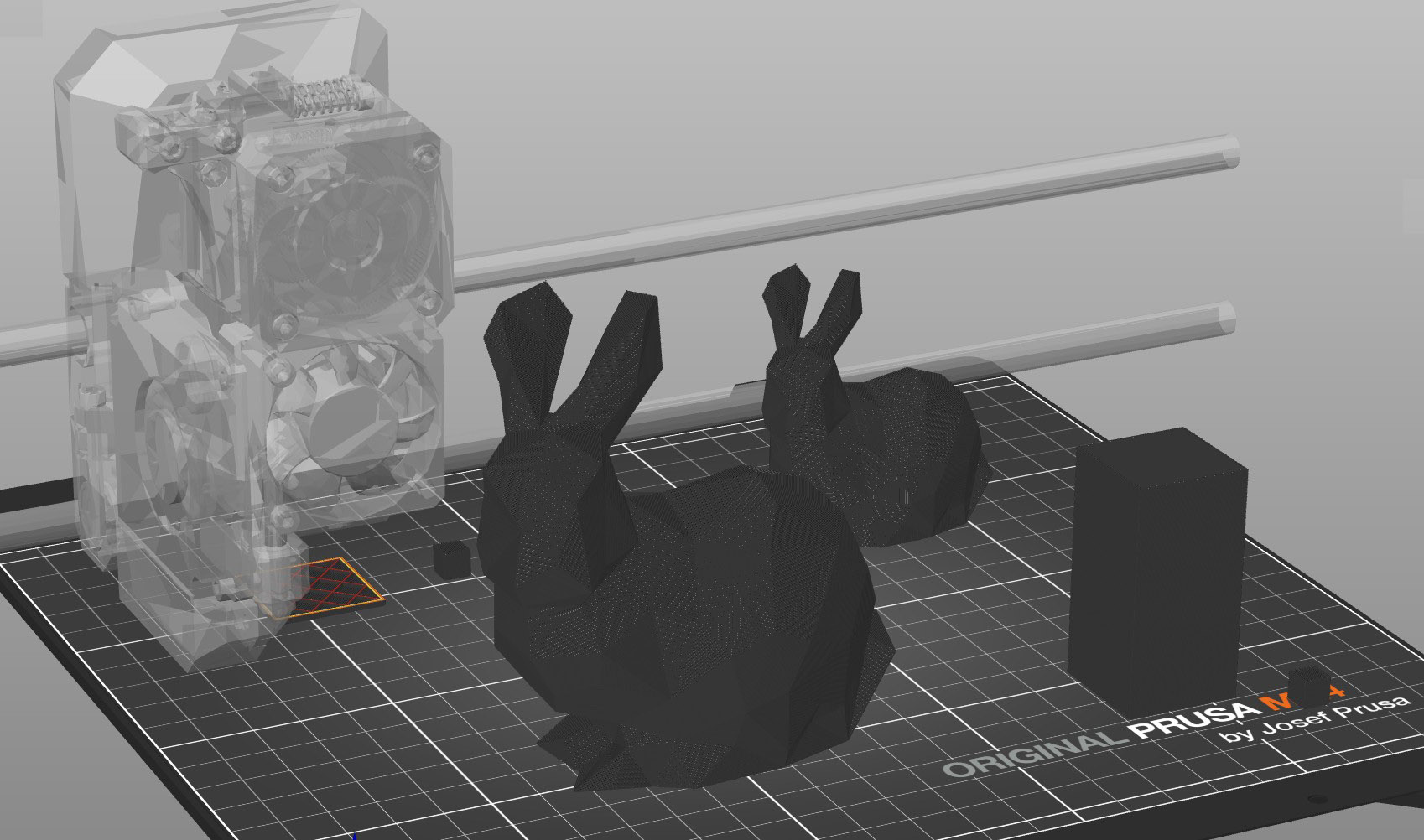}
    \end{subfigure}
    \caption{Standard 3D printing slice by slice and sequential 3D printing where objects are completed one by one shown in Prusa Slicer \cite{prusa-slicer-2025}. The ordering of objects for sequential printing is shown by numbers. Printer extruder and gantry must avoid peviously printed objects in the sequential case (printing of the last object is shown).}
    \label{fig:seq-print}
\end{figure*}

In this work, however, we go further, we deal with the task of the so-called {\em sequential 3D printing}, where we will not print all objects in slice-by-slice at one, but will complete the objects one after other, with individual objects being printed in a standard slice-by-slice manner. The problem is particularly challenging because the objects need to be arranged on the printing plate in such a way that the print head and other mechanical parts, such as the gantry on which the print head is mounted, avoid previously printed objects (see Figure \ref{fig:seq-print} for illustrations). Moreover, sequential printing does not only mean spatial arrangement of objects on a printing plate, but also determining the order in which the objects are printed.

%Navic sekvencni tisk neznamena jenom prostorove umisteni objektu, ale take urceni poradi jejich tisku.

Sequential printing has far-reaching significance for modern 3D printing, it can help to tackle the following challenges:

\begin{enumerate}[label=(\roman*)]
\item increasing the rebustness of the printing process to errors (in case of failure, we do not have to repeat the entire print, but only the unfinished objects)
\item maximize the printing speed (sequential printing eliminates frequent movements of the print head between objects)
\item minimize the number of time consuming color changes during multi-color printing, where a classic example is printing each object in a different color.
\end{enumerate}

\subsection{Related Work}
%V souvisejicim tematu rectangle packing, který ovsem predstavuje méně obecný problém, existuje řada výsledku.
%There are a number of results on the related topic of {\em rectangle packing}, which is a less general problem.

Abstract versions of the problem of object arrangement on a printing plate have been studied in the literature. Search-based algorithms have been developed for the related problem of {\em rectangle packing} \cite{DBLP:conf/socs/HuangK11,DBLP:journals/jair/HuangK13}. As shown in \cite{DBLP:conf/aips/Korf03}, rectangle packing is NP-hard, which means that object arrangement and scheduling for sequential printing is also NP-hard, since it is a more general problem.

%Jak je ukazano v , rectangle packing je NP-tezky, coz znamena, ze rozvrhovani pro sekvencni tisk rovnez, nebot se jedna o obecnejsi ulohu.

Genetic optimization has been used for packing 3D objects \cite{Ikonen1997AGA}. Various special algorithms have also been developed for box packing \cite{DBLP:conf/ijcai/LimY01} or algorithms for placing smaller objects in concave parts of other objects \cite{DBLP:journals/eor/EgebladNO07}. The most relevant to us are works that translate problems, whether rectangle packing or 3D packing, into other formalisms, such as those for {\em constraint programming}  (CSP) \cite{DBLP:books/daglib/0016622}, {\em linear programming} LP \cite{rader2010deterministic}, or {\em satisfiability modulo theories} (SMT) \cite{DBLP:reference/mc/BarrettT18} for which an off-the-shelf efficient solvers exist. Rectangle packing using CSP is descibed in \cite{DBLP:conf/aips/MoffittP06,DBLP:journals/anor/KorfMP10} and the application of SMT for the same problem is shown in \cite{Nikken2020}.

%-genetic \cite{Ikonen1997AGA} 3d 
%-rectanlge packing \cite{DBLP:conf/socs/HuangK11,DBLP:journals/jair/HuangK13}.

%-box pack \cite{DBLP:conf/ijcai/LimY01}
%-convex/concave \cite{DBLP:journals/eor/EgebladNO07}

%-SMT results \cite{Nikken2020}
%-CSP results \cite{DBLP:journals/anor/KorfMP10}, \cite{DBLP:conf/aips/MoffittP06}

%Abstraktni verze problemu umistovani objektu na plat byly v literature studovany. Algoritmy zalozene na prohledavani byly vyvinute pro souvisejici problem rectangle packing. Pro packing 3D objektu byla pouzita geneticka optimalizace. Byly vyvinute i ruzne specialni algoritmy pro box packing ci algoritmy umistujici mensi objekty do konkavnich casti jinych objektu. Nam nejblizsi jsou prace, ktere problemy at uz rectangle packing ci 3D packing, prevadeji do jineho formalismu, jako jsou pro constraint programming (CSP), linearni programovani, nebo SMT. 

\subsection{Contribution}

%The combinatorial nature of the problem suggests that it would be appropriate to consider using one of the formalisms and related solvers for combinatorial satisfiability and optimization, such as {\em constraint satisfaction} (CSP)  \cite{DBLP:books/daglib/0016622}, {\em linear programming} \cite{rader2010deterministic}, {\em satisfiability modulo theories} (SMT) \cite{DBLP:reference/mc/BarrettT18}.

The problem arranging and scheduling objects for sequential printing represents a significant challenge in terms of task modeling and combinatorial optimization. We first focused on providing a formal definition of the problem, as it seems that the problem has not yet been formally described in the literature. Based on the formal definition, we will create a simplified model that will be friendly for subsequent solving, i.e. for calculating the positions and the ordering of objects. 

%Z kombinatorické povahy problému vyplývá, že by bylo vhodné uvažovat o použití některého z formalismů a souvisejících řešičů pro kombinatorickou splnitelnost a optimalizaci, jako jsou formalisy constraint satisfaction (CSP) nebo satisfiability modulo theories (SMT).

%Nami navrzeny model vyjadruje otazky arranging and scheduling for sequential printing jako formuli linearni aritmetiky, kterou nasledne resime pomoci resice pro satisfiability modulo theories. Ukazujeme ze primocare namodelovani problemu pomoci formule linearni aritmetiky nestaci, nebot ta je v zpravidla prilis tezka na vyreseni. Proto jsme navrhli techniku reseni formule linearni aritmetiky modelujici problem sekvencniho tisku, ktera je inspirovana technikami counterexample guided abstraciton refinement (CEGAR). V experimentech ukazujeme, ze pouziti nasi CEGAR-inspirovane techniky predstavuje klicovou inovaci pro dosazeni efektivniho reseni problemu.

Our proposed model expresses the question of arranging and scheduling for sequential printing as a linear arithmetic formula \cite{DBLP:series/txtcs/KroeningS16}, which we then solve using an off-the-shelf solver for {\em satisfiability modulo theories}. We show that modeling the problem using a linear arithmetic formula is not enough, since it is impractically hart to solve. Therefore, we proposed a technique for solving a linear arithmetic formula modeling the sequential printing problem, which is inspired by {\em counterexample guided abstraction refinement} (CEGAR) \cite{DBLP:conf/cav/ClarkeGJLV00,DBLP:journals/jacm/ClarkeGJLV03} techniques. In experiments, we show that the use of our CEGAR-inspired technique represents a key innovation for achieving an efficient solution to the problem.

\section{Background}

%Je to otázka určení pozic a pořadí objektů tak, že objekty může v určeném pořadí na určených pozicích vytisknout 3D tiskárna postupně jeden po druhém. Na rozdíl od standardního 3D tisku, kdy jsou po jednotlivých vrtvách tištěny všechny objety najednou, jsou při sekvenčním tisku po jednotlivých vrstvách objekty dokončovány indivuduálně. V okamžiku tištění dalšího objektu jsou tedy dříve vytištěné objekty stále přítomné na tiskovém plátu. Je tedy nutné zajistit, aby při tisku dalšího objektu nedošlo ke kolizi tiskové hlavy a dalších mechanických částí 3D tiskárny s dříve vytištěnými objekty.

The problem of 3D object sequential arrangement is the task of determining object positions on the printing plate and the order of printing so that objects can be printed in the order one by one by a 3D printer. 

The problem of {\em 3D sequential object arrangement and scheduling} (inspired by the terminology for rectangle packaging, we will rather call the problem ``object packing and scheduling'' and denote SEQ-PACK+S) is a task of determining the positions and order of 3D objects so that the objects can be printed by the 3D printer in a determined order at the determined positions sequentially, one after the other. Unlike standard 3D printing, where all objects are printed at once in individual slices, in sequential printing, objects are completed individually. That is, when the next object is printed, the previously printed objects are still present on the printing plate. It is therefore necessary to ensure that the print head and other mechanical parts of the 3D printer do not collide with previously printed objects when printing the next object.

%Problem predstavuje vyznamnou vyzvu z hlediska modelovani úloh a kombinatoricke optimalizace. Zameřili jsme se nejprve na podání formální definice problému, neboť se zdá, že formálně problém v literatuře zatím popsán není. Na základě formální definice vytvoříme zjednodušený model, který bude přívětivý pro následné vyřešení, tedy pro vypočtení pozic a pořadí objektů.

% Na realne kartezske tiskarne je extruder object reprezentovan extruderem, tiskovou hlavou, gantry, kabely a dalsimi pohyblivymi castmi. Predpokladame, ze nemeni behem sveho pohybu tvar. Cili v tomto matematickem modelu napriklad abstrahujeme od ohybani kabelu. Nicmene, za model lze dale zobecnit, aby i toto bral v potaz.

While determining the printing order of objects to print is a discrete problem, determining object positions is inherently a continuous problem, which presents specific challenges.

%Zatimco urceni poradi tisku objektu je diskretni problem, urcovani pozic objektu je ve sve podstate spojity problem, coz predstavuje specificke vyzvy.

Let $\mathbb{R}^3$ be a three dimensional Euclidean space, a finite set of objects $\mathbb{O} = \{ O_1, O_2, ..., O_k\}$, where each object $O_i$ is a connected set $\{(x_i, y_i, z_i) \in O_i\} \subseteq \mathbb{R}^3$, in addition to this, there is an {\em extruder} object $E$, $\{(x_e, y_e, z_e) \in E\} \subseteq \mathbb{R}^3$ that represents a moving part of the printer that prints objects. On a real Cartesian 3D printer, the extruder object is represented by the extruder, print head, gantry, cables, and other moving parts. All objects including the extruder are expressed in the same coordinate system. We assume that the extruder does not change shape during its motion. For example, in this mathematical model we abstract from bending of cables. However, the model can be further generalized to take this into account. When printing extruder $e$ moves, that is, $E$ is translated to some position $(x_t,y_t,z_t) \in \mathbb{R}^3$ at time $t \in \mathbb{R}$. Hence extruder appears as a translated object $\{(x_e + x_t, y_e + y_t, z_e + z_t) \;|\; (x_e, y_e, z_e) \in E\}\}$ at time $t$. We can assume that $(x_t,y_t,z_t)$ changes smoothly as with real 3D printers happens but it is not important for further definitions.

%Standardni Fused deposition modeling (FDM) 3D tisk probíhá tak, že tištěné objekty jsou umístěny na tiskový plát, resp. jsou postupně po vrstvách na plátu tištěny. Plát můžeme matematicky modelovat jako podmnožinu roviny kolmé na osu z. Typicky je plát obdélníkový, někdy kruhový, vyjímečně jiného tvaru. Plát budeme značit jako P. Nad plátem se pomocí kartézské mechaniky pohybuje extruder.

We can mathematically model the plate as a subset of a plane perpendicular to the $z$-axis. Typically, the plate is rectangular, sometimes circular, exceptionally of a different shape. Mathematically, the plate will be a subset of 2D plane, $P_P \subseteq \mathbb{R}^2$. The extruder moves above the plate using printer mechanics. Any point relatively above the printing plate $P$ up to certain height is accessible by the extruder which together defines a {\em printing volume} $\mathbb{V} \subseteq \mathbb{R}^3$.

For simplicity, we will assume that all objects to be printed will fit into the printing volume $\mathbb{V}$, and we will also assume that the extruder can move slightly above the printing volume, which eliminates the need to worry about the height of objects in further definitions.

Common 3D printer mechanics such as {\em Cartesian} (bed-slinger) or CoreXY are covered by our definitions.

%Pro zjednodušení budeme předpokládat, že všechny objekty k tisku se vejdou do printing volume, navíc budeme předpokládat, že extruder se může přesouvat i mírně nad printing volume, což eliminuje nutnost starat se v dalších definicích o výšky objektů.

%Idelane matematicky muzeme na tisk objektu nahlizet tak, ze je nutne vytvorit kazdy bod tisteneho objektu, tedy se jej dotknout extruderem, presneji receno vybranym bodem extruderu, v nasem pripade bodem (0,0,0) extruderu. Bod (0,0,0) extruderu by v případě reálné 3D tiskárny odpovídal otvoru trysky.

From the mathematical point of view, we can look at printing an object in such a way that it is necessary to create every point of the printed object, that is, to touch every point of the printed object with the extruder, or more precisely, with a selected point of the extruder, in our case the point (0,0,0) of the extruder. The point (0,0,0) of the extruder would correspond to the nozzle opening.

%Extruder se tedy musí posunout na každý bod tištěného objektu. Jakmile je bod natištěn, musíme počítat s jeho přítomností a nesmíme do něj extruderem později narazit, tj. natištěný bod se nesmí ocitnout uvnitř posunutého extruderu.

%Při standardním tisku předpokldáváme, že objekty umístíme na plát a tiskneme je podle rostoucí souřadnice z po vrstvách. Za předpokladu, že z-souřadnice extruderu jsou kromě bodu (0,0,0) nezáporné (což odpovídá i reálné 3D tiskárně), nehrozí kolize s žádnou již natištěnou částí.

The extruder must therefore move to each point of the printed object. Once a point is printed, we must take into account its presence for future movements of the extruder, that is, the extruder must not collide with the point in the future. Mathematically, the printed point must never appear inside the translated extruder.

In standard printing, we assume that we place objects on the plate and print them according to the increasing $z$-coordinate in slices. Assuming that the $z$-coordinates of the extruder are non-negative except for the point (0,0,0) (which also corresponds to a real 3D printer), there is no risk of collision with any already printed part. However, this is no longer case in sequential 3D printing, when individual objects are completed one after other.

%V pripade sekvencniho tisku chceme urcit pozice objektů a pořadí objektů, aby při postupném tisku objektů v tomto určeném pořadí nikdy nedošlo ke kolizi s extruderu a dříve vytištěného objektu.

In the case of sequential printing, we want to determine the positions of the objects and their temporal ordering, so that when printing objects sequentially in this specified order, there is never a collision with the extruder and a previously printed object. We will call this requirement a sequential {\em non-colliding requirement}. Next, we need to ensure that all objects are placed on the printing plate, we will call this requirement the {\em printing plate requirement} (no part of an object extends beyond the printing plate) \footnote{Equivalently, the printing plate requirement can be expressed by the requirement that all objects are placed within the printing volume $V$. The way of expression we have chosen corresponds better with the proposed solution technique.}. And finally, we need to ensure that the extruder can be lifted vertically up and moved to print the next object, i.e. the extruder can be lowered back to the printing plate for printing the next object. We can simplify the last requirement slightly by requiring the extruder to be able to move freely vertically upwards from any top point of the object being printed. This allows both for initial accessing the volume for printing the object as well as leaving the object volume. We will call this requirement an {\em extruder traversability} requirement. Other variations of this requirement are also possible, and the method we developed is general enough to work with such generalizations.

%Nami vybrany zpusob vyjadreni lepe koresponduje s navrzenou resici technikou.

%  v pripade sekvencniho tisku chceme urcit pozice objektů a pořadí objektů, aby při postupném tisku objektů v tomto určeném pořadí nikdy nedošlo ke kolizi s extruderu a dříve vytištěného objektu.
% Dale musime zajistit, ze vsechny objekty jsou umistene na tiskovem platu, teto podmince budeme rikat printing plate requirement. A nakonec musime zajistit, ze se extruder po dokonceni tisku objektu muze zdvihnout a presunout na tisk dalsiho objektu, tj. nejdrive opet poklesnout až k tiskovemu platu. Poslední podmínkum můžeme mírně zjednodušit tak, že budeme pro extruder požadovat možnost volného vertikálního pohybu nahoru z libovolného vrchního bodu vytištěného objektu. Poslední podmínku můžeme mírně zjednodušit tak, že budeme pro extruder požadovat možnost volného vertikálního pohybu nahoru z libovolného vrchního bodu vytištěného objektu. Jiné varianty této podmínky jsou také možné a námi vyvinutá metoda je dostatečně obecná, aby s takovými zobecneními mohla pracovat.

% Formálně chceme určit pro každý objekt jeho pozici (x,y,z) a pořadí objektů, tedy permutaci

Formally, we need to determine for each object $O_i \in \mathbb{O}$ its position on the plate $(X_i,Y_i,Z_i) \in \mathbb{R}^3$ ($Z_i$ is determined by placing object vertically on the surface of the plate) and the order of the objects, i.e. the permutation $\pi: \{1,2,...,k\} \rightarrow \{1,2,...,k\}$, such that the {\em sequential non-colliding requirement}, the {\em printing plate requirement} hold, and {\em extruder traversability} hold.

We introduce object transforming functions: $\mathcal{P}: 2^{\mathbb{R}^3} \rightarrow 2^{\mathbb{R}^3}$, a function that places object onto the plate, $\mathcal{E}: 2^{\mathbb{R}^3} \rightarrow 2^{\mathbb{R}^3}$, a function that makes {\em extruder envelope} of a given object.

In addition to this, let $\mathit{()}^{xy}: 2^{\mathbb{R}^3} \rightarrow 2^{\mathbb{R}^2}$ be a {\em projection} of the given object onto the printing plate, and $\mathit{()}^{\top}: 2^{\mathbb{R}^3} \rightarrow 2^{\mathbb{R}^3}$ be an {\em extended top} of the printed object defined as follows:

%Printing plate is denoted $\mathbb{P_P}$.

%Let $\mathcal{P}(O_i)$ be a {\em printed object} corresponding to $O_i$, let $\mathit{EPO}_i$ be an {\em extruder envelope} of printed object $O_i$, let $\mathit{PO}_i^{xy}$ be a {\em projection} of the printed object onto the printing plate, and $\overline{\mathit{PO}}_i$ be an {\em extended top} of the printed object defined as follows:

\begin{itemize}
\item $\mathcal{P}(O_i) = \{(x_i + X_i, y_i + Y_i, z_i + Z_i)\;|\\
\;(x_i,y_i,z_i) \in O_i\}$

\item $\mathcal{E}(O_i) = \{(x_i + x_e, y_i + y_e, z_i + z_e)\;|\\
\;(x_i,y_i,z_i) \in O_i \wedge (x_e,y_e,z_e) \in E\}$

%\item $\mathit{EO}_i = \{(x_i + x_e, y_i + y_e, z_i + z_e)\;|\\
%\;(x_i,y_i,z_i) \in O_i \wedge (x_e,y_e,z_e) \in E\}$

\item $\mathit{O}_i^{xy} = \{(x,y)\;|\;(x,y,z) \in O_i\}$

\item $\mathit{O_i}^{\top} = \{(x,y,z)\;|\;z \geq {\max}_z \wedge (x,y,{\max}_z) \in O_i \}$ \\where ${\max}_z=\max\{z\;|\;(x,y,z) \in O_i\}$
\end{itemize}

Let us note that $\mathcal{E}(\mathcal{P}(O_i))$ and $\mathcal{E}(O_i)$ are equivalent to the Minkowski sum of $\mathcal{P}(O_i)$ and the extruder $E$ and $O_i$ and the extruder $E$ respectively.

The {\bf sequential non-colliding} requirement can be expressed as follows:

\begin{equation}
\label{con:sequential-noncolliding}
(\forall i,j=1,2,...,k) (\pi(i) < \pi(j) \Rightarrow \mathcal{P}(O_i) \cap \mathcal{E}(\mathcal{P}(O_j)) = \emptyset)
\end{equation}

Placement objects within the printing plate, i.e. the {\bf printing plate} requirement can be expressed as follows:

\begin{equation}
\label{con:on-plate}
(\forall i=1,2,...,k) (\mathcal{P}(O_i)^{xy} \subseteq \mathbb{P})
\end{equation}

The {\bf extruder traversability} requirement can be expressed as follows:

\begin{equation}
\label{con:extruder-traversability}
(\forall i,j=1,2,...,k) (\pi(i) < \pi(j) \Rightarrow \mathcal{P}(O_i) \cap \mathcal{P}(O_j)^{\top} = \emptyset)
\end{equation}

{\bf Optimality in SEQ-PACK+S}.The problem of object packing for sequential printing is a decision problem in its basic variant. Various notions of optimality in SEQ-PACK+S can be adopted.

%Tiskovy plát reálné 3d tiskarny je obvykle vyhřívaný pro dosažení přilnutí objektů k plátu, přičemž nejrovnoměrnější vyhřívání je směrem ke středu plátu zatímto směrem k okrajům plátu narůstají neregularity ve vyhřívání. Vzhledem k těmto fyzikálním vlastnostem je výhodné umísťovat tištěné objekty směrem ke středu tiskového plátu. Tuto preferenci zohledníme i našem návrhu účelové funkce. 

The printing plate of a real 3D printer is usually heated to achieve adhesion of objects to the plate, with the most uniform heating towards the center of the plate, while irregularities in heating increase towards the edges of the plate. Given these physical properties, it is advantageous to place printed objects towards the center of the printing plate. We will also take this preference into account in our design of the objective function. Let $C_{P} \in \mathbb{R}^2$ be a center of $P_P$.

Intuitivelly, we will try to shrink $P_P$ around  $C_{\mathbb{P}}$ so that placement of object satisfying the sequential printability requirements is still possible. Formally, let $C_{P}=(x_c,y_c)$ and $\sigma{P_P} = \{(x_c+\sigma(x - x_c), y_c+\sigma(y - y_c))\;|\;(x,y) \in P_P\}$. The printing plate requirement can be modified accordingly:

\begin{equation}
\label{con:on-plate-opt}
(\forall i=1,2,...,k) (\mathcal{P}(O_i)^{xy} \subseteq \sigma{P_P})
\end{equation}

We will define the optimization variant of SEQ-PACK+S as finding $\sigma \in (0,1]$ that is as small as possible and requirements \ref{con:sequential-noncolliding}, \ref{con:extruder-traversability}, and \ref{con:on-plate-opt} are satisfied.

\section{A Linear Arithmetic Model}
In this section we will develop a linear arithmetic formula that models the SEQ-PACK+S problem. The described model of SEQ-PACK+S will simplify the problem from 3D space to 2D that enables expressing it as satisfaction of a linear arithmetic formula and consequently decision procedures for linear arithmetic formulae can be used.

 Let us first state several useful proposition that our model will rely on, some of them without proof.

\begin{proposition}
Minkowski sum of convex polygons $P_A=(A_1,A_2,...A_{\alpha}) \subseteq \mathbb{R}^2$ and $P_B=(B_1,B_2,...,B_{\beta}) \subseteq \mathbb{R}^2$ is convex polygon $P_C=(C_1,C_2,...,C_{\gamma}) \subseteq \mathbb{R}^2$.
\end{proposition}

For two points $A,B \in \mathbb{R}^2$, let $L[A,B] \subseteq \mathbb{R}^2$ denote a line segment connecting points $A$ and $B$.

\begin{proposition}
\label{prop:polygon-outside-polygon}
Two polygons $P_A=(A_1,A_2,...A_{\alpha}) \subseteq \mathbb{R}^2$ and $P_B=(B_1,B_2,...,B_{\beta}) \subseteq \mathbb{R}^2$  do not share any point if and only if the following two requirements hold:
\begin{enumerate}[label=(\roman*)]
\item $(\forall a=1,2,...,\alpha)A_a \not \in P_B \wedge (\forall b=1,2,...,\beta)B_b \not \in P_A$
\item $(\forall a=1,2,...,\alpha;\;\forall b=1,2,...,\beta)\\L[A_a,A_{(a\bmod \alpha)+1}] \cap L[B_b,B_{(b\bmod \beta)+1}] = \emptyset$
\end{enumerate} 
\end{proposition}

%Tvrzeni predstavuje dava navod na efektivni detekci skutecnosti, ze se konvexni polygony neprekryvaji, nebot kazda cast podminky lze vyjadrit  jako formule lineární aritmetiky, přičemž pro formule lineární aritmetiky máme určité rozhodovací procedury.

The proposition provides a calculation for efficiently detecting the fact that convex polygons do not overlap, since each part of the requirement can be expressed as a {\em linear arithmetic formula}, and for linear arithmetic formulas we have certain decision procedures \cite{DBLP:series/txtcs/KroeningS16}.

%Intuitivně by se mohlo zdát, že stačí pro nepřekrytí konvexních polygonů podmínka (i), ovšem, jak ukazuje následující příklad, není to pravda.

\begin{figure}[h]
    \centering
    \includegraphics[trim={2.1cm 22.5cm 3.5cm 2.6cm},clip,width=0.75\textwidth]{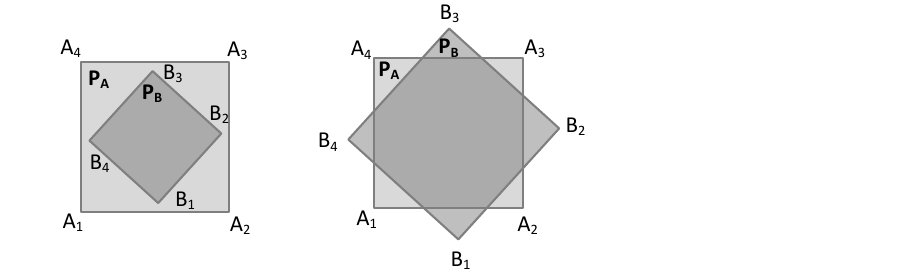}
    \vspace{-0.4cm}
    \caption{Two overlapping convex polygons $P_A=(A_1,A_2,A_3,A_4)$ and $P_B=(B_1,B_2,B_3,B_4)$. Left: Condition (i) is violated while requirement (ii) is satisfied, Right: Although requirement (i) is met, the polygons overlap. Condition (ii) is violated.}
    \label{fig:convex-nonoverlap}
\end{figure}

Intuitively, it might seem that requirement (i) is sufficient for convex polygons to not overlap, but as the example in Figure \ref{fig:convex-nonoverlap} shows, this is not true.

\begin{proof}
$\Rightarrow$: If $P_A$ and $P_B$ do not share any point then requirements (i) and (ii) since violation of any of them implies the existence of a common point.

$\Leftarrow$: By contradiction, suppose that there is a common point $C \in P_A \cap P_B$. Then we choose a direction and go from point C in the chosen direction. We must necessarily encounter an edge of either polygon $P_A$ or polygon $P_B$. Let the first encounter of an edge be without loss of generality edge $L[A_a,A_{(a \bmod \alpha)+1}]$ from polygon $P_A$ at point $C'$. For $C'$ it holds that it is both in $P_A$ and $P_B$ since $C'$ is the first encounter of a polygon edge. Then we continue along the edges of polygon $P_A$ (counter clock wise). If we go around the entire polygon $P_A$ back to point $C'$ and do not encounter any edge of polygon $P_B$, this means that we have not left polygon $P_B$ and therefore all points $A_1, A_2, ..., A_{\alpha}$ lie inside polygon $P_B$, which violates requirement (i). If while going around polygon $P_A$ we encounter an edge of polygon $P_B$, say at point $C''$, the intersection of the edge of polygon $P_A$ and the edge of polygon $P_B$ occurs at point $C''$, which violates requirement (ii). In both cases we have reached a contradiction.
\end{proof}

Notice that we do not need the convex property to show the proposition. Convex property will become useful as soon as we try to express the requirements (i) and (ii) computationally.

% Potom vyberme smer a vydejme se od bodu C ve vybranem smeru. Nutně musíme narazit na hranu buď polygonu A, nebo polygonu B. Nechť je to bez ujmy na narazime na hranu L polygonu A.
%Pak pokracujeme pohranach polygonu A counter clock wise. Pokud obejdeme cely polygon A zpet do bodu C a pritom nenarazime na zadnou hranu polygonu B, znamena to, ze jsme neopustili polygon B a tudiz vsechny body A+, Aě, ..., Ak lezi uvnitr polygonu B, coz porusuje podminku (i). Pokud pri obchazeni polygonu A narazime na hranu polygonu B, rekneme v bode C'', dochazi v bode C'' k pruseciku hrany polygonu A a hrany polygonu B, coz je poruseni podminky (ii). V obou pripadech jsme dosli ke sporu.

% Budeme rovnez potrebovat efektivne vyjadrit, ze jeden konvexni polygon lezi uvnitr druheho.

We will also need to effectively express that one convex polygon lies inside another. This can be described by the following proposition (without proof).

\begin{proposition}
\label{prop:polygon-inside-polygon}
For two convex polygons $P_A=(A_1,A_2,...A_{\alpha}) \subseteq \mathbb{R}^2$ and $P_B=(B_1,B_2,...,B_{\beta}) \subseteq \mathbb{R}^2$ it holds $P_A \subseteq P_B$ if and only if $(\forall a = 1,2,...,\alpha)A_a \subseteq P_B$.
\end{proposition}

%$L[A_i,A_{(i\bmod k)+1}]$
%$A_a.x$, $A_a.y$

Let $X_A,Y_A \in \mathbb{R}$ and $X_B, Y_B \in \mathbb{R}$ be decision variables of a linear arithmetic formula determining positions of polygons $P_A$ and $P_B$ respectively. Let $U_a = (A_{(a\bmod \alpha)+1} - A_a)$ be a direction vector of edge $L[A_a,A_{(a \bmod \alpha)+1}]$, let $N_a$ be a normal vector to $U_a$, that is $N_a =\begin{bmatrix}U_a.y  \\ -U_a.x \end{bmatrix}$. Then we can express that point $(B_b.x + X_B, B_b.y + Y_B)$ lies outside the half-plane given by the point $(A_a.x + X_A, A_a.y + Y_A)$ and the normal vector $N_a$ can be expressed by the floowing linear enequality.

%lezi mimo polorovinu danou bodem ax a normalovym vektorem N lze vyjadrit nasledujici linearni nerovnici

%$B_j.x, B_j.y$

\begin{equation}
\label{eq:point-outside-halfplane}
\left(\begin{bmatrix}B_b.x  \\ B_b.y \end{bmatrix} + \begin{bmatrix}X_B  \\ Y_B \end{bmatrix} - \begin{bmatrix} A_a.x \\ A_a.y \end{bmatrix} - \begin{bmatrix} X_A \\ Y_A \end{bmatrix} \right) \cdot \begin{bmatrix}U_a.y  \\ -U_a.x \end{bmatrix} > 0
\end{equation}

Let $X_i, Y_i \in \mathbb{R}$ be decision variables determining the position of object $O_i$. Let  {\em Point-outside-Halfplane} or $\mathit{PoH}(X_A, Y_A, A_a,$ $A_{(a\bmod \alpha)+1},X_B, Y_B,B_b)$ be a short-hand for constraint (\ref{eq:point-outside-halfplane}). Similarly, a constraint expressing that the point  $(B_b.x + X_B, B_b.y + Y_B)$ must lie inside a half-plane given by the point $(A_a.x + X_A, A_a.y + Y_A)$ and the normal vector $N_a$ will be denoted as {\em Point-inside-Halfplane} or $\mathit{PiH}(X_A, Y_A, A_a,$ $A_{(a\bmod \alpha)+1},X_B, Y_B,B_b)$. Mathematically, $\mathit{PiH}$ differs from $\mathit{PoH}$ only by the opposite inequality ($\leq0$) in \ref{eq:point-outside-halfplane}.

Let $V_b = (B_{(b\bmod \beta)+1} - B_b)$ be a direction vector of edge $L[B_b,B_{(b + \bmod \beta)+1}]$. The non-intersection between two line segments $L[A_a,A_{(a \bmod \alpha)+1}$ and $L[B_b,B_{(b \bmod \beta)+1}]$ that are not parallel can be expressed using the following constraints:

%neprotnuti mezi dvema useckami, ktere nejsou rovnobezne, lze vyjadrit pomoci nasledujicich podminek.

\begin{equation}
\label{eq:line-nonintersection-1}
\begin{bmatrix}A_a.x  \\ A_a.y \end{bmatrix} + \begin{bmatrix}X_i  \\ Y_i \end{bmatrix} + t \cdot U_a =
\begin{bmatrix}B_b.x  \\ B_b.y \end{bmatrix} + \begin{bmatrix}X_j  \\ Y_j \end{bmatrix} + t' \cdot V_b
\end{equation}

\begin{equation}
\label{eq:line-nonintersection-2}
t < 0 \vee t > 1 \vee t' < 0 \vee t' > 1
\end{equation}

% maji lokalni platnost v ramci uvedene podminky, jinymi slovy promenne vzdy zavadime cersve pro novou podminku.

Let us note that decision variables $t$ and $t'$ have local scope within the specified constraint only. In other words, we always introduce fresh variables $t$ and $t'$ for a new constraint.

Let  {\em Lines-not-Intersect} or $\mathit{LnI}(X_A, Y_A, A_i,A_{(i\bmod k)+1},$ $X_B,Y_B,B_j,B_{(j \bmod l)+1})$ be a short-hand for constraints (\ref{eq:line-nonintersection-1}) and (\ref{eq:line-nonintersection-2}). While (\ref{eq:line-nonintersection-1}) is a linear program as its individual linear equations (a linear equations for the $x$ and $y$ coordinate) are connected by the {\bf and} connective, constraint (\ref{eq:line-nonintersection-2}) is not due to the {\em or} connective, it is a general linear arithmetic formula.

Let $X_A,Y_A \in \mathbb{R}$ and $X_B, Y_B \in \mathbb{R}$ be decision variables. For two convex polygons $P_A=(A_1,A_2,...A_{\alpha}) \subseteq \mathbb{R}^2$ and $P_B=(B_1,B_2,...,B_{\beta}) \subseteq \mathbb{R}^2$, where $P_A$ is placed at a coordinate $(X_A,Y_A)$ and $P_B$ is placed at a coordinate $(X_B,Y_B)$ we can express that polygons $P_A$ and $P_B$ do not overlap using the following linear arithmetic formula:

\begin{equation}
\label{eq:point-outside-polygon1}
\begin{split}
 \bigvee_{a = 1,2,...,\alpha; b = 1,2,...,\beta}
 \mathit{PoH}(X_A, Y_A, A_a,A_{(a\bmod \alpha)+1},\\X_B,Y_B,B_b)
\end{split}
\end{equation}

\begin{equation}
\label{eq:point-outside-polygon2}
\begin{split}
 \bigvee_{a = 1,2,...,\alpha; b = 1,2,...,\beta}
 \mathit{PoH}(X_B, Y_B, B_b,B_{(b\bmod \beta)+1},\\X_A, Y_A,A_a)
\end{split}
\end{equation}

\begin{equation}
\label{eq:polygon-lines-nonintersection}
\begin{split}
 & (\forall a = 1,2,...,\alpha)(\forall b = 1,2,...,\beta) \\
 & \mathit{LnI}(X_A, Y_A, A_a,A_{(a\bmod \alpha)+1},X_B, Y_B,B_b,B_{(b\bmod \beta)+1}) \wedge \\
 & \mathit{LnI}(X_B, Y_B, B_b,B_{(b\bmod \beta)+1},X_A, Y_A,A_a,A_{(a\bmod \alpha)+1})
\end{split}
\end{equation}

%Vsimneme si, ze podminky jsou brany jako konjunkce, ale jejich interni boolovska struktura konjunkci neni, takze celkove se nejedna o linearni program ale o obencou formuli linearni aritmetiky.

We will handle the first part of the formula, that is constraints (\ref{eq:point-outside-polygon1}) and (\ref{eq:point-outside-polygon2}) and the other part (\ref{eq:polygon-lines-nonintersection}) in a slightly different way in our solver.

We will use a shorthand {\em Points-outside-Polygon} or $\mathit{PoP}(X_A, Y_A, P_A, X_B, Y_B, P_B)$ for (\ref{eq:point-outside-polygon1}) and (\ref{eq:point-outside-polygon2}) and a shorthand {\em Polygon-Lines-not-Intersect} or $\mathit{PLnI}(X_A,Y_A,$ $P_A, X_B, Y_B, P_B)$.

%S podminkami A a B a s podminkou C budeme zachazet trochu jinym zpusobem.

Let us note that constraints (\ref{eq:point-outside-polygon1}), (\ref{eq:point-outside-polygon2}), and (\ref{eq:polygon-lines-nonintersection}) are treated as conjunction, but their internal Boolean structure is not a conjunction, so altogether this is not a linear program but a general linear arithmetic formula. Constraints (\ref{eq:point-outside-polygon1}) and (\ref{eq:point-outside-polygon2}) are negations of the standard expression stating that a point lies inside all half-planes formed by the edges of the convex polygon.

We will also need a constraint for expressing that a convex polygon, say $P_A$, is subsumed by another convex polygon, say $P_B$. Due to Proposition \ref{prop:polygon-inside-polygon} we can express $P_A \subseteq P_B$ as follows:

\begin{equation}
\label{eq:polygon-inside-polygon}
\begin{split}
 \bigwedge_{a = 1,2,...,\alpha; b = 1,2,...,\beta}
 \mathit{PiH}(X_B, Y_B, B_b,B_{(b\bmod \beta)+1},\\X_A, Y_A,A_a)
\end{split}
\end{equation}

This constraint is a conjunction of linear inequalities, that is it forms a linear program. The shorthand for constraint \ref{eq:polygon-inside-polygon} will be {\em Polygon-inside-Polygon} or $\mathit{PiP}(X_A, Y_A,P_A,X_B, Y_B,P_B)$.

We introduce decision variables $T_i \in \mathbb{R}$ determining the time at which respective objects $O_i$ are printed. $T_i$ variables will be used to determine the permutation $\pi$ of objects for sequential printing. Let $\mathcal{C}(O_i^xy)$ be a convex polygon (with finite number of edges) such that $O_i^{xy} \subseteq \mathcal{C}(O_i^{xy})$. Practically, $\mathcal{C}$ can be implemented as a convex hull of a projection of object $O_i$ onto the plate.

%Pomoci nasledujicich podminek budeme aproximovat podminky na sekvencni tisk.

\begin{figure}[h]
    \centering
    \includegraphics[trim={2.1cm 14.0cm 6.5cm 2.6cm},clip,width=0.35\textwidth]{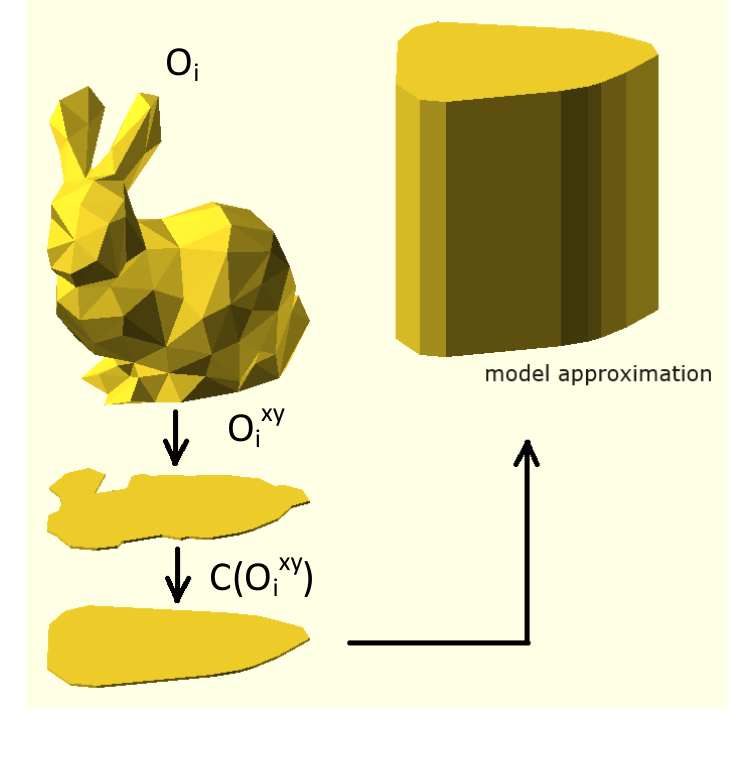}
    \vspace{-0.8cm}
    \caption{Illustration of object simplification used in modeling the requirements for sequential printing (SEQ-PACK+S) as a linear arithmetic formula.}
    \label{fig:model-approximation}
\end{figure}

Using the following constraints, we will approximate the requirements for sequential object packing and scheduling, that is, the sequential non-colliding requirement (\ref{con:sequential-noncolliding}) and the printing plate requirement (\ref{con:on-plate}).

\begin{equation}
\label{eq:seq-constraint}
\begin{split}
(\forall i=1,2,...,k; & j=1,2,...,k; i \neq j) \\
T_i < T_j \Rightarrow \; & (\mathit{PoP}(X_i, Y_i, \mathcal{C}(O_i^{xy}), X_j, Y_j, \mathcal{C}(\mathcal{E}(O_j)^{xy})) \\
                             \wedge \; & \mathit{PoP}( X_j, Y_j, \mathcal{C}(\mathcal{E}(O_j)^{xy}), X_i, Y_i, \mathcal{C}(O_i^{xy})) \\
                             \wedge \; & \mathit{PLnI}(X_i,Y_i, \mathcal{C}(O_i^{xy}), X_j, Y_j, \mathcal{C}(\mathcal{E}(O_j)^{xy}))
\end{split}
\end{equation}

The printing plate requirement can be expressed using constraint \ref{eq:polygon-inside-polygon} assuming that the printing plate $P$ is represented by a convex polygon which usually is the case as most real Cartesian 3D printers use rectangular printing plates. Let $P_P = (P_1,P_2,...,P_{\psi})$ be a convex polygon representing the printing plate $\mathbb{P}$. We assume that $P$ is fixed so naturally it is placed at position $(0,0)$ and all other objects will the positioned with respect to the position of $P_P$.

Extruder traversability constraint is automatically ensured in our model.

\begin{equation}
\label{eq:plate-constraint}
\begin{split}
(\forall i=1,2,...,k) \\
& \mathit{PiP}(X_i,Y_i,\mathcal{C}(O_i^{xy}),0,0,P_P)
\end{split}
\end{equation}

\begin{figure}[h]
    \centering
    \includegraphics[trim={2.1cm 22.5cm 3.5cm 2.5cm},clip,width=0.75\textwidth]{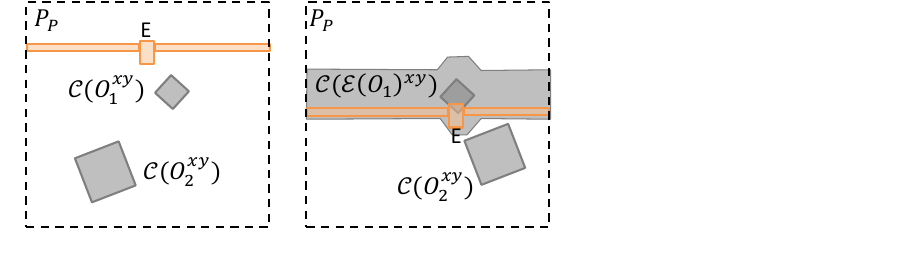}
    \vspace{-0.8cm}
    \caption{Illustration of the idea behind linear arithmetic formula modeling SEQ-PACK+S. In the setting where $O_1$ is printed after $O_2$ (right), the $xy$-projection of extruder $E$ and other moving parts of the printer moves only within $\mathcal{C}(\mathcal{E}(O_1)^{xy})$, hence ensuring non-overlapping between $\mathcal{C}(O_2^{xy})$ and $\mathcal{C}(\mathcal{E}(O_1)^{xy})$ as required by constraint (\ref{eq:seq-constraint}) ensures correct sequential printing.}
    \label{fig:idea-illustration}
\end{figure}

\section{A CEGAR Inspired Algorithm}

\begin{algorithm}[t!]
\begin{footnotesize}
\SetKwBlock{NRICL}{Solve-CEGAR-SEQ$(P_P, \{O_1,O_2,...,O_k\})$}{end} \NRICL{
    $(X_1,Y_1,T_1),...,(X_k,Y_k,T_k)) \gets ((\bot,\bot,\bot),...,(\bot,\bot,\bot))$\\
    $\mathcal{F} \gets []$ \\
    \For{each $i,j \in \{1,2,...,k\} \wedge i \neq j$}{
        $\mathcal{F} \gets \mathcal{F} \cup \{T_i  + \epsilon_T < T_j \vee T_j  + \epsilon_T < T_i\}$\\
        $\mathcal{F} \gets \mathcal{F} \cup \{T_i < T_j \Rightarrow \; (\mathit{PoP}(X_i, Y_i, \mathcal{C}(O_i^{xy}),
                                                                                                               X_j, Y_j, \mathcal{C}(\mathcal{E}(O_j)^{xy})) 
                                                          \wedge \; \mathit{PoP}( X_j, Y_j, \mathcal{C}(\mathcal{E}(O_j)^{xy}),
                                                                                           X_i, Y_i, \mathcal{C}(O_i^{xy})))\}$
    }
    $\sigma_{0} \gets 0$\\
    $\sigma_{+} \gets 1$\\    
    \While {$\sigma_{+} - \sigma_{0} > \epsilon_{XY}$} {
    	$\sigma \gets (\sigma_{+} + \sigma_{0}) / 2$ \\
    	$\mathit{answer} \gets$ Solve-CEGAR-SEQ-Bounded ($\mathcal{F},\sigma,P_P,\{O_1,O_2,...,O_k\})$\\
    	\If {$\mathit{answer} \neq \mathit{UNSAT}$}{
    		{\bf let} $(((X_1,Y_1,T_1),...,(X_k,Y_k,T_k)),\mathcal{F}) = \mathit{answer}$\\
    	    	$\sigma_{+} \gets \sigma$
    	}
    	\Else{
    		$\sigma_{0} \gets \sigma$
    	}
     }
     \Return{$((X_1,Y_1),...,(X_k,Y_k))$}
}

\SetKwBlock{NRICL}{Solve-CEGAR-SEQ-Bounded($\mathcal{F},\sigma,P_P,\{O_1,O_2,...,O_k\}$)}{end} \NRICL{
	$\Phi \gets []$\\
	\For{each $i \in \{1,2,...,k\}$}{
	    $\Phi \gets \Phi \cup \{\mathit{PiP}(X_i,Y_i,\mathcal{C}(O_i^{xy}),0,0,\sigma P_P)\}$
	}
	\While{$\mathit{TRUE}$}{
 	  $\mathit{answer} \gets$ Solve-SMT$(\mathcal{F},\Phi)$\\	
	  \If{$\mathit{answer} \neq \mathit{UNSAT}$}{
	    {\bf let} $((X_1,Y_1,T_1),...,(X_k,Y_k,T_k)) = \mathit{answer}$\\
	    \For{each $i,j \in \{1,2,...,k\} \wedge i \neq j$} {        
		{\bf let} $(A_1,A_2,...,A_{\alpha}) = \mathcal{C}(O_i^{xy})$ \\
		{\bf let} $(B_1,B_2,...,B_{\beta}) =  \mathcal{E}(\mathcal{C}(O_j^{xy}))$ \\
		$\mathit{refined} \gets \mathit{FALSE}$\\
		\For{each $a \in \{1,2,...,\alpha\}$}{
		  	\For{each $b \in \{1,2,...,\beta\}$}{
		  	    \If{$T_i < T_j \wedge$ Lines-Intersect$(X_i, Y_i, A_a,$\\
		  	                            $A_{(a\bmod \alpha)+1},
		  	                             X_j, Y_j,B_b,B_{(b\bmod \beta)+1})$}{
                           $\mathcal{F} \gets \mathcal{F} \cup \{T_i < T_j \Rightarrow
                                                                                                 \mathit{LnI}(X_i, Y_i, A_a,A_{(a\bmod \alpha)+1},$ \\
                                                                                                 $X_j, Y_j,B_b,B_{(b\bmod \beta)+1})\}$\\
                           $\mathit{refined} \gets \mathit{TRUE}$\\
		  	    }
		  	}
		}
		\If{$\neg \mathit{refined}$}{
		    \Return {$(((X_1,Y_1,T_1),...,(X_k,Y_k,T_k)),\mathcal{F})$}		
		}
	    }
       }
       \Else{
         \Return{$\mathit{UNSAT}$}
       }
    }
}

\caption{CEGAR-SEQ: Sequential object packing solving via CEGAR-inspired satisfaction of a linear arithmetics formula.}
\label{alg:CEGAR-SEQ}
\end{footnotesize}
\end{algorithm}

%Sice lze problem SEQ-PACK+S vyresit pouhym zavedenim popsanych omezeni a vyresenim omezeni pomoci off-the-shelf resice pro satisfiability modulo theories (SMT), ale takovy postup nepovazujeme za efektivni. V jine souvisejici uloze multi-agentniho hledani cest, ktera byla take resena zfromulovanim omezeni a jejich vyresenim pomoci off-the-shelf solverem, se ukazal jako efektivnejsi postup reseni inspirovany counterexample guided abstraction refinement (CEGAR). Nase predbezne experimenty rovnez ukazaly, ze i v uloze SEQ-PACK+S bude CEGAR-inspirovany postup vykonnejsi nez prime vyreseni omezeni resicem.

Although the suggested model of the SEQ-PACK+S problem can be solved simply by introducing the described constraints and solving the constraints using an off-the-shelf solver for {\em satisfiability modulo theories} (SMT) \cite{DBLP:series/faia/336,DBLP:journals/jacm/NieuwenhuisOT06}, we do not consider such an approach to be efficient. In another related problem where collision avoidance plays an important role, in multi-agent pathfinding problen (MAPF) \cite{DBLP:journals/ai/AndreychukYSAS22,DBLP:conf/iros/Surynek23}, which was also solved by formulating constraints and solving them using an off-the-shelf solver, a so called {\em counterexample guided abstraction refinement} (CEGAR) \cite{DBLP:conf/cav/ClarkeGJLV00,DBLP:journals/jacm/ClarkeGJLV03} inspired solving proved to be more efficient. Our preliminary experiments showed that in the SEQ-PACK+S problem too, the CEGAR-inspired approach will be more efficient than directly solving all the constraints in one shot.

\begin{figure*}[t]
    \centering
    \begin{subfigure}{0.49\textwidth}
       \includegraphics[trim={2.5cm 22.0cm 10.5cm 2.5cm},clip,width=1.0\textwidth]{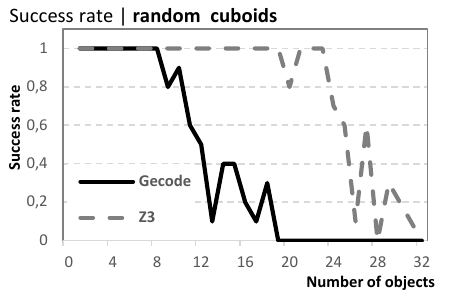}
    \end{subfigure}
    \begin{subfigure}{0.49\textwidth}
       \includegraphics[trim={2.5cm 22.0cm 10.5cm 2.5cm},clip,width=1.0\textwidth]{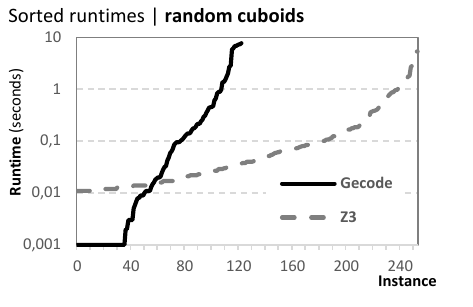}
    \end{subfigure}
    \caption{Comparison of solving SEQ-PACK+S by the Gecode solver and the z3 solver on random cuboids. The right part shows cactus plots of runtimes (lower plot is better).}
    \label{expr:solvers}
\end{figure*}

An algorithm for solving an instance of the approximation of SEQ-PACK+S as a linear arithmetic formula by a CEGAR-inspired approach is shown using pseudo-code as Algorithm \ref{alg:CEGAR-SEQ}, we call the algorithm CEGAR-SEQ.

The basic idea of the algorithm is to solve an abstract variant of the instance and refine it by adding constraints if necessary. The initial abstraction (formula $\mathcal{F}$) is constructed by introducing constraint \ref{eq:seq-constraint} from which the {\em Polygon-Lines-not-Intersect} part is omitted (line 6). In addition to this, temporal ordering for object priting over $T_i$ variables is introduced (line 5).

%Dalsi postup je takovy, ze se pokusime vyresit abstrakci (relaxaci) problemu pomoci off-the-shelf SMT resice pro formule linearni aritmetiky. Jestlize abstrakce resitelna neni, nemuze byt resitelny ani puvodni problem, kde by byly vsechny omezeni pritomne od zacatku. Jestlize abstrakce resitelna je, musime proceduralne zkontrolovat, zda se nektere hrany polygonu neprotinaji, tj. zda neni porusana cast Polygon-Lines-not-Intersect. Pritom je treba brat ohled na casove poradi objektu, nebot pozdejsi objekt se jevi zvetseny pomoci Minkowske sumy o objekt extruderu. Jestlize zadne protnuti hran polygonu nenajdeme, znamena to, ze jsme uspesne nasli reseni problemu. Jestlize ale protnuti hran polygonu najdeme, je potreba problem tedy formuli refinovat, tj. pridat porusenou podminku a vyslednou formuli vyresit znova. Vyhoda tohoto CEGAR-inspirovaneho postupu je, ze problem zpravidla vyresime drive, nez dojde k pridani vsech omezeni, coz zpravidla vede k uspore celkoveho casu behu.

The next step is to try to solve the abstraction (relaxation) $\mathcal{F}$ using an off-the-shelf SMT solver for linear arithmetic formulas (line 23). If $\mathcal{F}$ is not solvable, the original problem, where all the constraints were present from the beginning, cannot be solvable either (line 40).

If $\mathcal{F}$ is solvable (line 24), we must procedurally check whether some of the polygon edges corresponding to input objects $O_i$
do intersect (lines 30-32), i.e. whether the Polygon-Lines-not-Intersect part of constraint \ref{eq:seq-constraint} is violated. In doing so, we must take into account the teporal ordering of objects given by $T_i$ variables, because the later object appears to be augmented by the extruder object using the Minkowski sum (the later polygon appears as $\mathcal{E}(\mathcal{C}(O_j^{xy}))$).

If we do not find any intersections of the polygon edges, this means that we have successfully found a solution to the instance that can be returned (line 38). However, if we do find intersections of the polygon edges, we need to refine $\mathcal{F}$, i.e. add the violated constraint to $\mathcal{F}$ (lines 34-36) and solve the resulting formula again.

We assume that Solve-SMT($\mathcal{F},\Phi$) represents an incremental SMT solver for linear arithmetics formulae. The first parameter $\mathcal{F}$ represents the refined linear arithmetic formula and $\Phi$ represents assumptions expressing printing plate constraint \ref{eq:plate-constraint}, that is the SMT solver tries to find a satisfying assignment for $\mathcal{F} \wedge \Phi$ while learning is performed only with respect to $\mathcal{F}$.

%Cely CEGAR-SEQ algoritmus je rozdelen do dvou procedur A a B, ktere spolecne realizuji optimalizaci vzhledem k velikosti tiskoveho platu. Zatimco procedura B predstavuje rozhodovaci variantu pro danou velikost tiskoveho platu, procedura A provadi binarni vyhledavni co nejmensi velikosti platu, kam se rozvrhovane objekty jeste vejdou.

The CEGAR-SEQ algorithm is divided into two procedures Solve-CEGAR-SEQ and Solve-CEGAR-SEQ-Bounded, which together perform optimization with respect to the size of the printing plate represented by polygon $P_P$. Procedure Solve-CEGAR-SEQ-Bounded represents a yes/no decision whether sequential packing is possible for a given size of the printing plate, that is for plate $\sigma P_P$. Procedure Solve-CEGAR-SEQ performs binary search for the smallest possible plate size (lines 9-16), that is smallest parameter $\sigma$ is searched.

Let us note that we use the constant $\epsilon_{XY}$ which determines the granularity of the binary search that can be se arbitrarily small. Similarly we use the constant $\epsilon_{T}$ white determines minimum time interval between objects.

The advantage of this CEGAR-inspired approach is that we usually solve the problem before all the constraints are added, which usually leads to a saving in overall runtime.

\begin{proposition}
The CEGAR-SEQ algorithm finds an optimal solution with respect the the size of the printing plate for the approximation of a given SEQ-PACK+S instance $(P_P,\{O_1,O_2,...,O_k\})$.
\end{proposition}

PLnI is not violated often so it is a good candidate for the CEGAR-inspired refinement.

\section{Experimental Evaluation}

CEGAR-SEQ has been written in C++ and integrated as part of Prusa Slicer 2.9.1 \cite{prusa-slicer-2025}, an open-source slicing software for 3D printers. The Z3 Theorem Prover \cite{10.5555/1792734.1792766}, an SMT solver, has been used for solving the linear arithmetic model within CEGAR-SEQ. We have performed extensive testing of the CEGAR-SEQ algorithm and present some of the results in this section. All experiments were performed on a system with CPU AMD Ryzen 7 2700 3.2GHz, 32GB RAM, running Kubuntu Linux 22. All experimental data presented in this paper will be available on \texttt{github.com/surynek/cegar-seq} to support reproducibility of results.

\subsection{Comparison of Solvers}

%Jelikoz v soucasnosti chybi alternativni srovnatelny rozvrhovac pro sekvencni tisk, rozhodli jsme se kompetitivni srovnani udelat alespon na urovni resicu. Porovnali jsme reseni problemu SEQ-PACK+S v resicich z3 a v resici Gecode, coz je resic CSP.

Since there is currently no comparable solver for sequential printing, we decided to conduct a competitive comparison at least at the solver level. We compared solving of the SEQ-PACK+S problem with the Z3 solver and with the Gecode solver \cite{gecode2006}, which is a CSP solver. Hence we compared solving SEQ-PACK+S in two different paradigms, SMT and CSP.

The experimental setup consisted of a printing plate of size 250 $\times$ 210 \footnote{Sizes can be assumed in mm that corresponds to a real 3D printer.} and random cuboids whose length, width and height were integer and chosen randomly from a uniform distribution in the interval [8,64], the number of cuboids ranges from 1 to 32. For each number of cuboids, 10 random instances were generated. The timeout was set to 8 seconds.

Cuboids do not need the PLnI constraints (\ref{eq:polygon-lines-nonintersection}), hence are easier to schedule (extruder $E$ must be cuboid as well). Since the Gecode solver is based on different principles than the SMT solver, we had to modify the model, i.e. the formula modeling the problem. The main difference is that Gecode uses variables with a finite domain, so we used a domain that divides the printing plate by millimeters. Using cuboids allowed us to simplify constraints in model for Gecode as well, only inequalities similarly as in rectangle packing were used.

Results are shown in Figure \ref{expr:solvers} which shows clear dominance of the SMT paradigm for the given problem. Additionally, we note that the linear arithmetic formula uses variables with a rational domain, so the resulting solution from the Z3 solver is more accurate.

%Navic podotkneme, ze formule linearni aritmetiky pouziva promenne s racionalni domenou, takze vysledne reseni od resice z3 je presnejsi.

%Jelikoz je resic Gecode zalozen na odlisnych pricipech nez SMT resice, museli jsme model, tedy formuli modelujici problem upravit. Hlavni odlisnosti je, ze Gecode pouziva promenne s konecnou domenou, pouzili jsme tedy  pro umistovani objektu domenu rozdelenou po mm. Pro zjednoduseni jsme rozvrhovali pouze cuboidy, takze resic nepouzival komplikovana omezeni , ale pouze nerovnosti.

%Setup: printing plate of size 250 $\times$ 210 \footnote{Sizes can be assumed in mm that corresponds to a real 3D printer.}, random cuboids random cuboids whose length, width and height were chosen randomly from a uniform distribution in the interval [8,64], the number of objects ranges from 1 to 32, timeout 8 seconds. Cuboids do not need the PLnI constraints - easier to solve (extruder must be cuboid as well).

\subsection{Benefit of CEGAR-style Refinements}

%Otestovali jsme take vyznam CEGAR-inspirovaneho pristupu pro reseni formule linearni aritmetiky modelujici problem. Porovnali jsme CEGAR-SEQ s variantou, kdy jsou vsechna omezeni pridana najednou ve snazivem stylu.

\begin{figure*}[t]
    \centering
    \begin{subfigure}{0.30\textwidth}
       \includegraphics[trim={0.5cm 0.5cm 0.5cm 0.5cm},clip,width=1.0\textwidth]{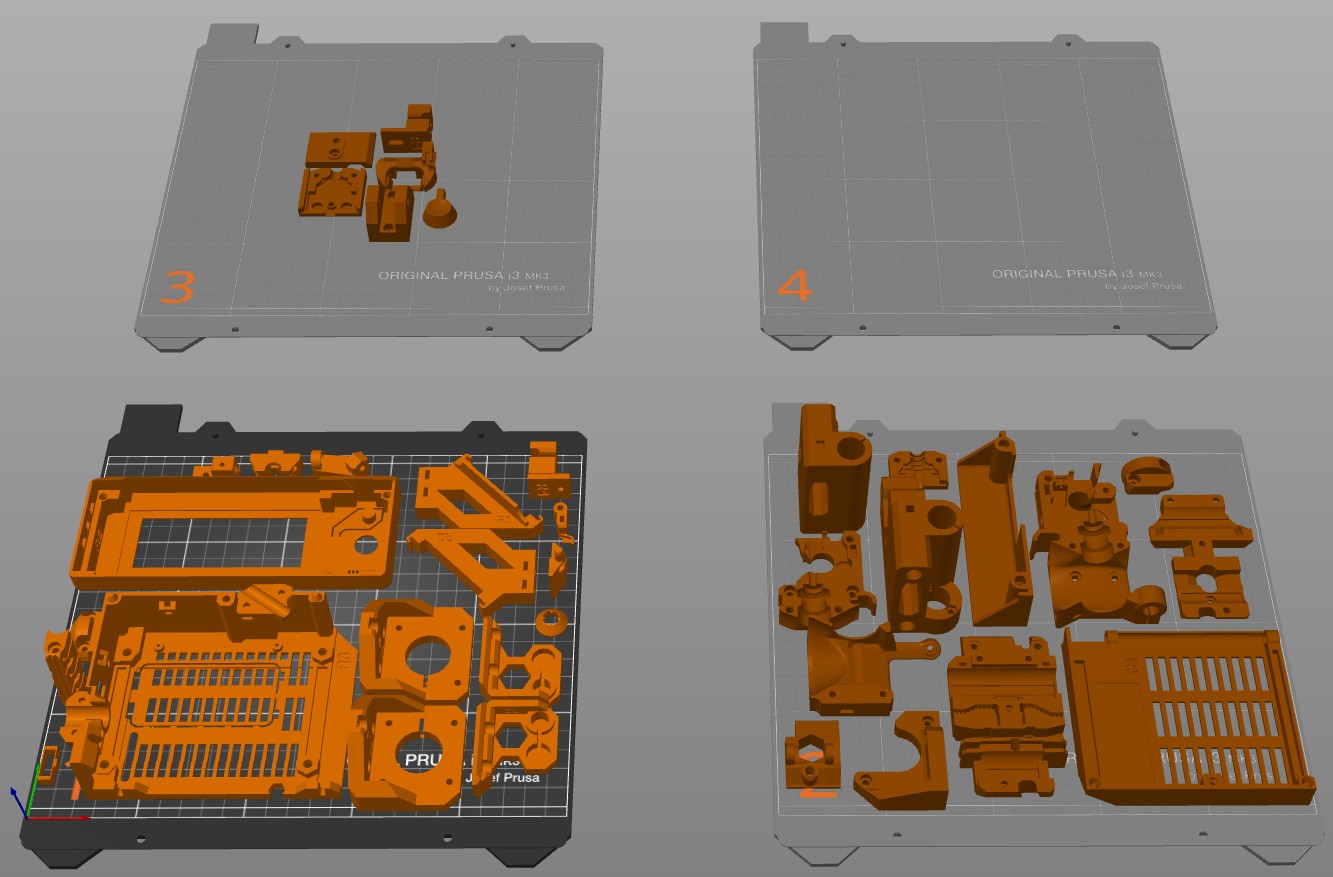}
    \end{subfigure}
    \begin{subfigure}{0.36\textwidth}
       \includegraphics[trim={2.5cm 22.0cm 10.5cm 2.5cm},clip,width=1.0\textwidth]{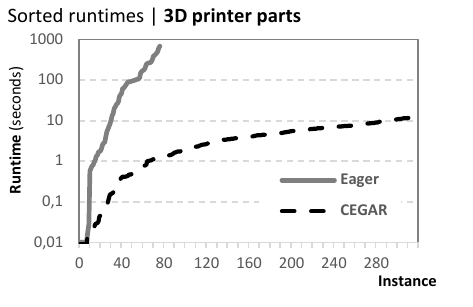}
    \end{subfigure}
    \begin{subfigure}{0.32\textwidth}
       \includegraphics[trim={0.5cm 0.5cm 0.5cm 0.5cm},clip,width=1.0\textwidth]{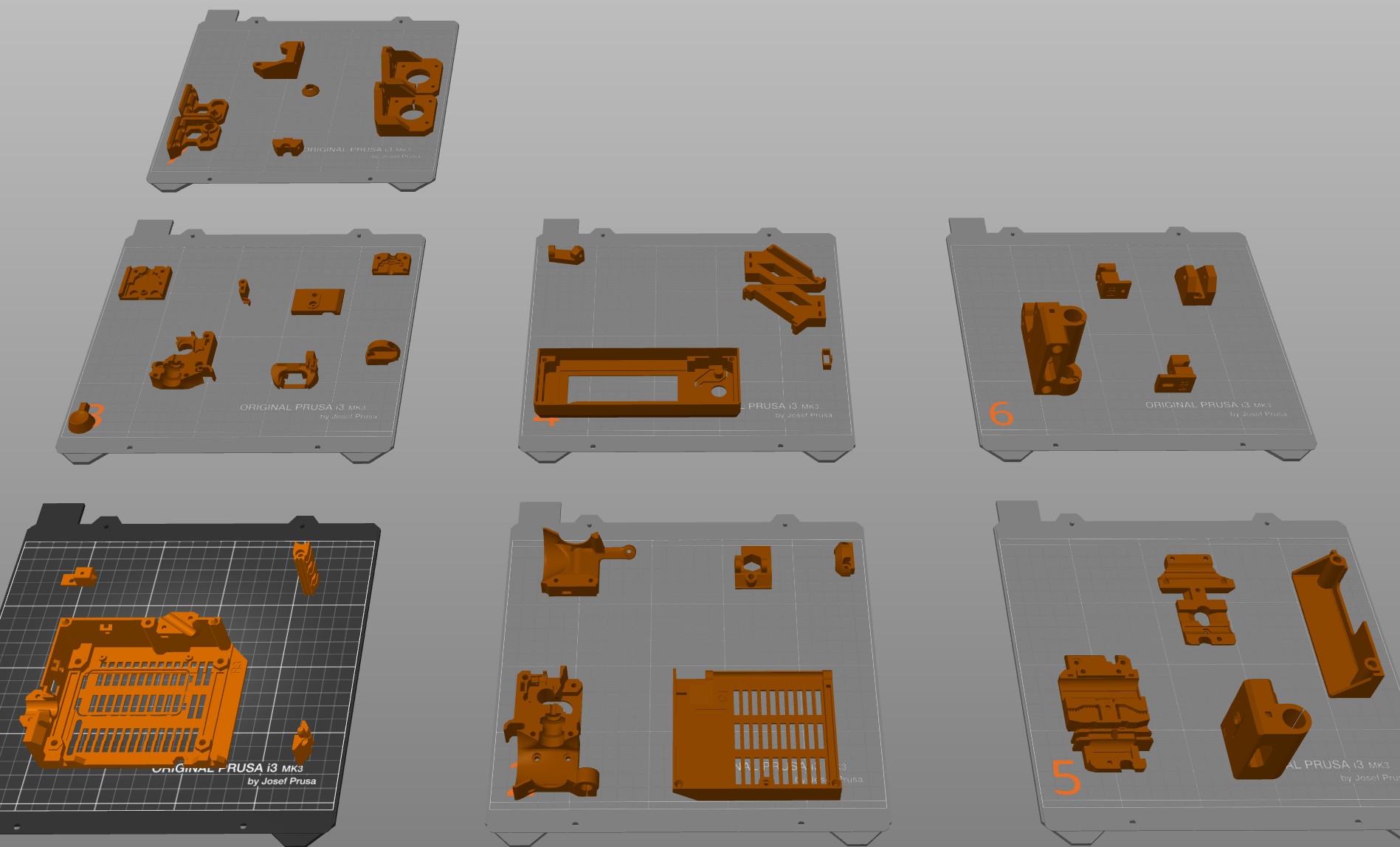}
    \end{subfigure}
    \caption{The effect of CEGAR-style refinements in contrast to eager encoding on many complex objects. Left: printable parts for a 3D printer (placed on 3 printing plates). Right: parts scheduled for sequential printing (7 printing plates used).}
    \label{expr:CEGAR}
\end{figure*}

We also tested the benefit of the CEGAR-inspired approach for solving the linear arithmetic formula modeling the problem. We compared CEGAR-SEQ with a variant where all constraints are added at once in an eager style.

As a testing case we use sequential 3D printing of complex objects - 3D printable parts for a 3D printer \footnote{The benchmark consists of 3D printable parts for the Original Prusa MK3S printer.}, the benchmark consists of 34 diverse objects, the timeout was set to 1000 seconds. We tried to solve SEQ-PACK+S for the increasing number of objects ranging from 1 up to 32. For each number of objects we selected random objects from the benchmark set (objects can be repeated), 10 instances per number of objects were solved.

Solving takes place across multiple printing plates - if objects do not fit onto a plate then remaining objects are scheduled on a fresh printing plate. Optimization with respect to placing as many as possible objects onto the plate towards the center was done.

Results are shown in Figure \ref{expr:CEGAR}. The results convincingly show that using CEGAR-style refinement represent a key technique for performance of the method as it provides significantly better performance than the eager variant.

\section{Conclusion}

%Navrhli jsme formalizaci problemu rozmistovani a rozvrhovani objektu pro sekvencni 3D tisku a zaroven jsme problem zjednodusili a zformulovali jej jako formuli linearni aritmetiky. Nas novy algoritmus CEGAR-SEQ pro reseni tohoto problemu, ktery jsme vybodovali s vyuzitim SMT resice z3, je zalozeny na myslence counterexample guided abstraction refinement (CEGAR). Formuli modelujici problem totiz neresime najednou, ale obtizne pripady kolizi mezi objekty testujeme proceduralne a pripadne provadime refinement formule. Pouziti CEGAR-inspirovane techniky reseni formule modelujici problem sekvencniho tisku se ukazalo jako klicove.

%Do budouci prace planujeme zpresnit linearne aritmeticky model, ktery by bral vice v uvahu jak se mení xy-průmět se zmenou z-souradnice objektu. Zároveň se chceme zabýbat možností otáčení objektů.

We proposed a formalization of the object arrangement (packing) and scheduling problem for sequential 3D printing denoted SEQ-PACK+S and at the same time we formulated a model of SEQ-PACK+S based on a linear arithmetic formula. Our new algorithm CEGAR-SEQ for solving SEQ-PACK+S, which we have built on top of the Z3 SMT solver, is based on the idea of counterexample guided abstraction refinement (CEGAR). We do not solve the formula modeling the problem in a straightforward way in one shot, but we test difficult cases of collisions between objects procedurally and refine the formula based on the result of the test. Using CEGAR-inspired techniques for solving the formula modeling the sequential printing problem turned out to be key innovation ensuring acceptable performance.

In our future work, we plan to improve our linear arithmetic model, which would take more into account how the $xy$-projection of objects changes with the $z$-coordinate. At the same time, we want to deal with the possibility of rotating objects.

%\section*{References}

\bibliographystyle{IEEEtran}
\bibliography{references}

\begin{thebibliography}{10}
\providecommand{\url}[1]{#1}
\csname url@samestyle\endcsname
\providecommand{\newblock}{\relax}
\providecommand{\bibinfo}[2]{#2}
\providecommand{\BIBentrySTDinterwordspacing}{\spaceskip=0pt\relax}
\providecommand{\BIBentryALTinterwordstretchfactor}{4}
\providecommand{\BIBentryALTinterwordspacing}{\spaceskip=\fontdimen2\font plus
\BIBentryALTinterwordstretchfactor\fontdimen3\font minus
  \fontdimen4\font\relax}
\providecommand{\BIBforeignlanguage}[2]{{%
\expandafter\ifx\csname l@#1\endcsname\relax
\typeout{** WARNING: IEEEtran.bst: No hyphenation pattern has been}%
\typeout{** loaded for the language `#1'. Using the pattern for}%
\typeout{** the default language instead.}%
\else
\language=\csname l@#1\endcsname
\fi
#2}}
\providecommand{\BIBdecl}{\relax}
\BIBdecl

\bibitem{DBLP:journals/cad/GaoZRRCWWSZZ15}
W.~Gao, Y.~Zhang, D.~Ramanujan, K.~Ramani, Y.~Chen, C.~B. Williams, C.~C.~L.
  Wang, Y.~C. Shin, S.~Zhang, and P.~D. Zavattieri, ``The status, challenges,
  and future of additive manufacturing in engineering,'' \emph{Comput. Aided
  Des.}, vol.~69, pp. 65--89, 2015.

\bibitem{multi-objective-packing2014}
S.~Wu, M.~Kay, R.~King, A.~Vila-parrish, and D.~Warsing, ``Multi-objective
  optimization of 3d packing problem in additive manufacturing,'' \emph{IIE
  Annual Conference and Expo 2014}, pp. 1485--1494, 01 2014.

\bibitem{DBLP:conf/ki/EdelkampW15}
S.~Edelkamp and P.~Wichern, ``Packing irregular-shaped objects for 3d
  printing,'' in \emph{{KI} 2015: Advances in Artificial Intelligence - 38th
  Annual German Conference on AI, Dresden, Germany, September 21-25, 2015,
  Proceedings}, ser. Lecture Notes in Computer Science, vol. 9324.\hskip 1em
  plus 0.5em minus 0.4em\relax Springer, 2015, pp. 45--58.

\bibitem{prusa-slicer-2025}
{The Prusa Slicer Team}, ``Prusa slicer 2.9.1,'' 2025, available from
  \texttt{https://github.com/prusa3d/PrusaSlicer}.

\bibitem{DBLP:conf/socs/HuangK11}
E.~Huang and R.~E. Korf, ``Optimal packing of high-precision rectangles,'' in
  \emph{Proceedings of the Fourth Annual Symposium on Combinatorial Search,
  {SOCS} 2011, Castell de Cardona, Barcelona, Spain, July 15.16, 2011}.\hskip
  1em plus 0.5em minus 0.4em\relax {AAAI} Press, 2011, pp. 195--196.

\bibitem{DBLP:journals/jair/HuangK13}
------, ``Optimal rectangle packing: An absolute placement approach,'' \emph{J.
  Artif. Intell. Res.}, vol.~46, pp. 47--87, 2013.

\bibitem{DBLP:conf/aips/Korf03}
R.~E. Korf, ``Optimal rectangle packing: Initial results,'' in
  \emph{Proceedings of the Thirteenth International Conference on Automated
  Planning and Scheduling {(ICAPS} 2003), June 9-13, 2003, Trento,
  Italy}.\hskip 1em plus 0.5em minus 0.4em\relax {AAAI}, 2003, pp. 287--295.

\bibitem{Ikonen1997AGA}
I.~Ikonen, W.~E. Biles, A.~Kumar, J.~C. Wissel, and R.~K. Ragade, ``A genetic
  algorithm for packing three-dimensional non-convex objects having cavities
  and holes,'' in \emph{International Conference on Genetic Algorithms}, 1997.

\bibitem{DBLP:conf/ijcai/LimY01}
A.~Lim and W.~Ying, ``A new method for the three dimensional container packing
  problem,'' in \emph{Proceedings of the Seventeenth International Joint
  Conference on Artificial Intelligence, {IJCAI} 2001, Seattle, Washington,
  USA, August 4-10, 2001}, B.~Nebel, Ed.\hskip 1em plus 0.5em minus 0.4em\relax
  Morgan Kaufmann, 2001, pp. 342--350.

\bibitem{DBLP:journals/eor/EgebladNO07}
J.~Egeblad, B.~K. Nielsen, and A.~Odgaard, ``Fast neighborhood search for two-
  and three-dimensional nesting problems,'' \emph{Eur. J. Oper. Res.}, vol.
  183, no.~3, pp. 1249--1266, 2007.

\bibitem{DBLP:books/daglib/0016622}
R.~Dechter, \emph{Constraint Processing}.\hskip 1em plus 0.5em minus
  0.4em\relax Elsevier Morgan Kaufmann, 2003.

\bibitem{rader2010deterministic}
D.~Rader, \emph{Deterministic Operations Research: Models and Methods in Linear
  Optimization}.\hskip 1em plus 0.5em minus 0.4em\relax Wiley, 2010.

\bibitem{DBLP:reference/mc/BarrettT18}
C.~W. Barrett and C.~Tinelli, ``Satisfiability modulo theories,'' in
  \emph{Handbook of Model Checking}, E.~M. Clarke, T.~A. Henzinger, H.~Veith,
  and R.~Bloem, Eds.\hskip 1em plus 0.5em minus 0.4em\relax Springer, 2018, pp.
  305--343.

\bibitem{DBLP:conf/aips/MoffittP06}
M.~D. Moffitt and M.~E. Pollack, ``Optimal rectangle packing: {A} meta-csp
  approach,'' in \emph{Proceedings of the Sixteenth International Conference on
  Automated Planning and Scheduling, {ICAPS} 2006, Cumbria, UK}.\hskip 1em plus
  0.5em minus 0.4em\relax {AAAI}, 2006, pp. 93--102.

\bibitem{DBLP:journals/anor/KorfMP10}
R.~E. Korf, M.~D. Moffitt, and M.~E. Pollack, ``Optimal rectangle packing,''
  \emph{Ann. Oper. Res.}, vol. 179, no.~1, pp. 261--295, 2010.

\bibitem{Nikken2020}
T.~Nikken, ``Satisfiability modulo theories based packing of scalable
  rectangles for real-time layout generation,'' \emph{Master Thesis, Radboud
  University Nijmegen}, pp. 1--53, 2020.

\bibitem{DBLP:series/txtcs/KroeningS16}
D.~Kroening and O.~Strichman, \emph{Decision Procedures - An Algorithmic Point
  of View, Second Edition}, ser. Texts in Theoretical Computer Science. An
  {EATCS} Series.\hskip 1em plus 0.5em minus 0.4em\relax Springer, 2016.

\bibitem{DBLP:conf/cav/ClarkeGJLV00}
E.~M. Clarke, O.~Grumberg, S.~Jha, Y.~Lu, and H.~Veith, ``Counterexample-guided
  abstraction refinement,'' in \emph{Computer Aided Verification, 12th
  International Conference, {CAV} 2000, Chicago, IL, USA, July 15-19, 2000,
  Proceedings}, ser. Lecture Notes in Computer Science, E.~A. Emerson and A.~P.
  Sistla, Eds., vol. 1855.\hskip 1em plus 0.5em minus 0.4em\relax Springer,
  2000, pp. 154--169.

\bibitem{DBLP:journals/jacm/ClarkeGJLV03}
------, ``Counterexample-guided abstraction refinement for symbolic model
  checking,'' \emph{J. {ACM}}, vol.~50, no.~5, pp. 752--794, 2003.

\bibitem{DBLP:series/faia/336}
A.~Biere, M.~Heule, H.~van Maaren, and T.~Walsh, Eds., \emph{Handbook of
  Satisfiability - Second Edition}, ser. Frontiers in Artificial Intelligence
  and Applications.\hskip 1em plus 0.5em minus 0.4em\relax {IOS} Press, 2021,
  vol. 336.

\bibitem{DBLP:journals/jacm/NieuwenhuisOT06}
R.~Nieuwenhuis, A.~Oliveras, and C.~Tinelli, ``Solving {SAT} and {SAT} modulo
  theories: From an abstract davis--putnam--logemann--loveland procedure to
  dpll(\emph{T}),'' \emph{J. {ACM}}, vol.~53, no.~6, pp. 937--977, 2006.

\bibitem{DBLP:journals/ai/AndreychukYSAS22}
A.~Andreychuk, K.~S. Yakovlev, P.~Surynek, D.~Atzmon, and R.~Stern,
  ``Multi-agent pathfinding with continuous time,'' \emph{Artif. Intell.}, vol.
  305, p. 103662, 2022.

\bibitem{DBLP:conf/iros/Surynek23}
P.~Surynek, ``Counterexample guided abstraction refinement with non-refined
  abstractions for multi-goal multi-robot path planning,'' in \emph{{IROS}},
  2023, pp. 7341--7347.

\bibitem{10.5555/1792734.1792766}
L.~De~Moura and N.~Bj\o{}rner, ``Z3: an efficient smt solver,'' in
  \emph{Proceedings of the Theory and Practice of Software, 14th International
  Conference on Tools and Algorithms for the Construction and Analysis of
  Systems}.\hskip 1em plus 0.5em minus 0.4em\relax Berlin, Heidelberg:
  Springer-Verlag, 2008, p. 337–340.

\bibitem{gecode2006}
{The Gecode Team}, ``Gecode: Generic constraint development environment,''
  2006, available from \texttt{http://www.gecode.org}.

\end{thebibliography}

%\vfill

\end{document}